\documentclass[prd,12pt]{revtex4}
\usepackage{amsmath,amssymb} % Enhanced mathematics
\usepackage[dvips]{graphicx} % Include figure files
\usepackage{graphics}
\usepackage{graphicx}
\usepackage{epsfig}
\usepackage[english]{babel}
\newcommand\nn{\nonumber}
\newcommand\ba{\begin{eqnarray}}
\newcommand\ea{\end{eqnarray}}
\newcommand\ee{\end{equation}}
\newcommand\be{\begin{equation}}

\newcommand{\brm}[1]{\left| #1 \right|}
\newcommand{\GeV}{~\mbox{GeV}}

\begin{document}

\title{Peripherical processes 2 $\to$ 3 and 2 $\to$ 4 in QED and QCD in $p(\bar p)p$ high energy collisions}

\author{A.~I.~Ahmadov}
\email{ahmadov@theor.jinr.ru}
\affiliation{Joint Institute for Nuclear Research, Dubna, Russia}
\affiliation{Institute of Physics, Azerbaijan National Academy of Sciences, Baku, Azerbaijan}

\author{Yu.~M.~Bystritskiy}
\email{bystr@theor.jinr.ru}
\affiliation{Joint Institute for Nuclear Research, Dubna, Russia}

\author{E.~A.~Kuraev}
\email{kuraev@theor.jinr.ru}
\affiliation{Joint Institute for Nuclear Research, Dubna, Russia}

\date{\today}

\begin{abstract}
Differential cross section of processes with high energy $p(\bar p)p$ collisions in frames of QED:
creation of scalar, pseudoscalar and lepton pair - are considered in Weizs\"acker - Williams
approximation. In frames of QCD processes with conversion of initial proton (antiproton) to
fermionic jets accompanied with one gluon jet as well as the state of two gluons and quark-antiquark
pair (with out rapidity gap) are considered in frames of effective Reggion action of theory of
Lipatov. Process of creation of a Higgs boson accompanied with two fermionic jets is considered.
The azimuthal correlation in process of two gluon jet separated by rapidity gap is investigated.
Effects of gluon reggeization are taken into account.
Some distributions are illustrated by numerical calculations.
\end{abstract}

\maketitle

%----------------------------------------------------
\section{Introduction}
\label{Introduction}
%----------------------------------------------------
Motivation of this paper is the construction of realistic formulae and the
estimation of the cross section with creation of two jets in the
proton (anti-proton) fragmentation regions and one or two additional jets in
multi-regge kinematics. Application of QCD methods to describe the peripherical
processes in high energy proton (antiproton)- proton scattering is based on
the proof of gluon reggeization phenomenon, which was done in papers ~\cite{Kuraev:1976ge} in
1973-1976 years.

For this purpose we use the effective Regge action ~\cite{Lipatov:1995pn,Antonov:2004hh}
of conversion of two reggeized gluons $R$ to some set of real particles $P,Q$-to
one and two gluons separated by rapidity gap, two gluon or quark-anti quark pair,
without rapidity gap and a scalar (Higgs) meson.

Construction of paper is as follows. After short review of QED processes in Section ~\ref{SectionQEDProcesses},
we consider processes of a single and two gluon production and production in Section ~\ref{SectionQCD1Processes}
and lepton pair in Section ~\ref{SectionQCD2Processes} , assuming the absence of rapidity gap between the couple
of particles created in pionization region. Measuring these processes provide the
possibility to check the $RRP,RRPP,RRq\bar{q}$ vertices of a effective Regge action
theory ~\cite{Antonov:2004hh}.
In Section ~\ref{Section5} the azimuthal correlation between two gluon jets separated by some rapidity gap is considered. In Section ~\ref{Sectionjjqq} the production of quark-antiquark is considered.
In Section ~\ref{Higgs} the Higgs boson production is considered.
In Conclusion we discuss the main topics of our approach and give the results of
numerical calculations.

%----------------------------------------------------
\section{QED Processes}
\label{SectionQEDProcesses}
%----------------------------------------------------

In the early seventies of the last century the processes of creation of some set of particles  were
intensively studied \cite{Baier:1980kx,Budnev:1974de}.
There was considered a different mechanism of pair production
in electron-positron collisions. The relevant formulae can in principle be applied
to proton-proton (antiproton) collisions.
Production os some set of particles in pionization region at high energy $p(\bar {p})p$ collision
\ba
p(\bar {p}) +p \to p(\bar {p}) + p +F
\ea
\ba
\biggl(\frac{s_1 d\sigma}{d s_1}\biggr)^{pp \to ppF} = \biggl(\frac{\alpha}{2\pi}\biggr)^2
\ln^2\biggl(\frac{s}{M_p^2}\biggr) \cdot f\biggl(\frac{s_1}{s}\biggr) \cdot \sigma_{tot}^{\gamma\gamma \to F}(s_1)
\biggl(1+O\biggl(\frac{1}{\ln s/{M_p^2}}\biggr)\biggr), \nn \\
f(z) = (2+z)^2 \ln\frac{1}{z} - 2(1-z)(3+z).
\ea
with $s_1$ - invariant mass square of the produced set of particles $F$.

\begin{figure}
\includegraphics[width=0.9\textwidth]{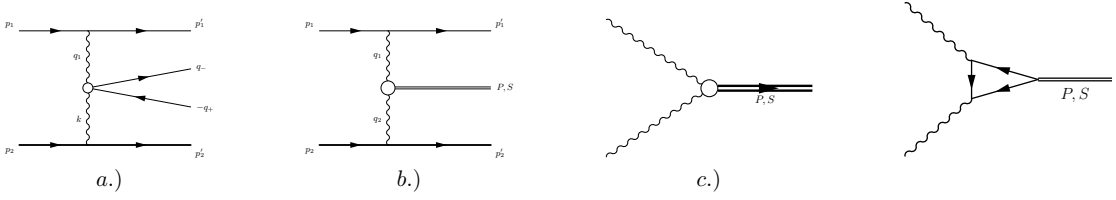}
\caption{Feynman diagrams for creation of 2 jet (a), 1 jet (b) by two reggeized gluons;
(c) --- creation of $P$, $S$-mesons by two reggeized gluons
\label{Fig1}}
\end{figure}

In the case when the lepton pair is created outside the fragmentation regions of protons the process cross
section (see Fig.~\ref{Fig1}, a)
\ba
p(p_1)+p(\bar{p})(p_2) \to p(p_1')+p(\bar{p})(p_2')+\mu^+(q_+)+\mu^-(q_-),
\ea
has the form
\ba
d\sigma ^{p \bar {p} \to l \bar {l} p \bar {p}}= \frac{2\alpha^4}{\pi}\frac{d^2q_1 d^2q_2 d^2 k_1 d x}{\pi^3}
\frac{d\beta_1}{\beta_1}\frac{\vec{q}_1^2\vec{q}_2^2}{(\vec {q}_1^2 +M^2 \beta_1^2)^2(\vec {q}_2^2+M^2 \alpha^2)^2}\cdot F;
\ea
where $M$ is the proton mass,
\ba
s\alpha = \frac{-c}{\beta_1 x(1-x)}, \,\,0<x=\frac{\beta_2}{\beta_1}<1; \,\,\beta_1 \ll 1,\\
c=m^2+\vec{q}_2^2+\vec{q}_1^2x +2\vec{q}_1\vec{q}_2x,\\
c_1=m^2+(\vec{k}_2-\vec{q}_2)^2+\vec{q}_1^2x+2\vec{q}_1(\vec{q}_2-\vec{k}_2)x;
\ea
\ba
\vec{q}_1^2\vec{q}_2^2F=\frac{\vec{q}_2^2\vec{q}_1^2}{cc_1}-\frac{x \bar x}{c^2c_1^2}[(\vec{q}_1^2+2\vec{q}_1 \vec{q}_2)
(\vec{q}_2^2-2\vec{k}_2\vec{q}_2)+2(\vec{q}_2\vec{q}_1)(m^2+\vec{k}_2^2)]^2.
\ea
Here $m$ is the lepton mass,
$x_1=1-\beta_1 \approx 1$, $-\vec{q}_1$ is the energy fraction of the scattered proton and its momentum
transversal to the initial proton direction $\vec{p}_1$ (center of mass of initial particles implied);
$1+\alpha \approx 1;$ \,\,$\vec{q}_2$ are similar quantities for the scattered proton (anti-proton); \,\,
$x\beta_1 +\frac{m^2+\vec{k}_1^2}{s\beta_1 x},$ $-\vec{k}_1$ and
$(1-x)\beta_1 +\frac{m^2+\vec{k}_2^2}{s\beta_1(1-x)},$
$\vec{k}_2=\vec{q}_1-\vec{q}_2-\vec{k}_1$ are the corresponding quantities for negative and
positive charged leptons from the pair created; $m$ is the mass of the created particle.

For two-photon processes with creation of pseudo-scalar and scalar particle we use the corresponding
subprocess $\gamma(q_1,\mu)+\gamma(q_2,\nu) \to P (S)$ (see Fig.~\ref{Fig1}, b, c)
with matrix elements described in terms
of triangle Feynman loop diagrams with quarks as internal fermions:
\ba
M^{\gamma\gamma P}=\frac{2\alpha N_Pg_p}{\pi m_q}(q_1e_1q_2e_2)I_P,
(q_1e_1q_2e_2)=\epsilon^{\alpha\beta\gamma\sigma}q_{1\alpha}
e_{1\beta}q_{2\gamma}e_{2\sigma}; \nn \\
M^{\gamma\gamma S}=\frac{2\alpha N_Sg_S}{\pi m_q}[q_1q_2)(e_1e_2)-(e_1q_2)(e_2q_1)]I_S,
\ea
where $e_{1,2}(q_{1,2})$ are the polarization vectors of photons, $N_{P,S}$ are the color factors
\ba
N_P=N_c(\frac{4}{9}-\frac{1}{9})=1; \nn \\
N_S=N_c(\frac{4}{9}+\frac{1}{9})=\frac{5}{3}.
\ea
Performing the loop momentum integration we obtain
\ba
I_{P,S}=\int\limits_0^1 d x\int\limits_0^1\frac{y dy }{d_{P,S}}(1,1-4y^2x(1-x)), \nn \\
d_{P,S}=1-y^2x(1-x)\frac{M^2_{P,S}}{m_q^2}-y(1-y)[x\frac{q_1^2}{m_q^2}+(1-x)\frac{q_2^2}{m_q^2}],
\ea
$M_{P,S},m_q$ are the masses of produced particles and quark mass. We can use the Goldberger--Treiman
relation $g_P/m_q=1/F_{\pi}$, with $F_{\pi}=93 MeV$ being the decay constant of charged pion, and a
similar relation $g_S/m_q=1/F_{\sigma}, F_{\sigma} \approx F_\pi$.

When inserting these matrix elements into the matrix element of process $2\to 3$ the combination is
used $M^{\gamma\gamma F}(e_1 \to p_1,e_2\to p_2)/s =m^{\gamma\gamma F}$. We obtain
\ba
m^{\gamma\gamma P}=\frac{\alpha N_P}{\pi F_\pi}[\vec{q}_1,\vec{q}_2]_z I_P; \nn \\
m^{\gamma\gamma S}=\frac{\alpha N_S}{\pi F_\sigma}(\vec{q}_1,\vec{q}_2)I_S,
\ea
where we consider the four-momenta of virtual photons to be essentially transversal two-
component Euclidean vectors $\vec{p}_1 \vec{q}_{1,2}=0$, $q_{1,2}^2=-\vec{q}_{1,2}^2<0$.

The cross sections of the processes of single meson production in the pionization region are
\ba
d\sigma^{pp\to pp P}=\frac{2\alpha^4}{\pi}\frac{d\beta_1}{\beta_1}dN_1 dN_2 C_P \sin^2\theta; \nn \\
d\sigma^{pp\to pp S}=\frac{2\alpha^4}{\pi}\frac{d\beta_1}{\beta_1}dN_1 dN_2 C_S \cos^2\theta,
\ea
where $\theta$ -- is the azimuthal angle between two-dimensional vectors $\vec{q}_1,\vec{q}_2$,
\ba
C_P=|\frac{N_P}{F_\pi}I_P|^2;C_S=|\frac{N_S}{F_\sigma}I_S|^2;
\ea
and Weizs\"acker--Williams (WW) enhanced factors
\ba
dN_1=\frac{\vec{q}_1^2 d^2\vec{q}_1}{(\vec{q}_1^2+m_p^2\beta_1^2)^2}, \nn \\
dN_2=\frac{\vec{q}_2^2 d^2\vec{q}_2}{(\vec{q}_2^2+m_p^2\alpha_2^2)^2}, \nn \\
\brm{s\alpha_2\beta_1}=M_{P,S}^2+(\vec{q}_1+\vec{q}_2)^2.
\ea
We use the expression of the squared 4-vectors of momenta transferred to a lepton pair:
\ba
q_1^2\approx -(\vec{q}_1^2+m_p^2\beta_1^2); \nn \\
q_2^2\approx -(\vec{q}_2^2+m_p^2\alpha_2^2).
\ea
These factors, being integrated, produce the "large logarithmic" factors
\ba
\frac{1}{\pi}\int\limits_{0}^{Q^2} d N_1=\ln\frac{q^2}{m_p^2\beta_1^2}-1, \,\,\,m_p^2 \ll Q^2 \ll s.
\ea

Cross section of production os some set of particles in pionization region
at high energy $p(\bar {p})p$ collision
\ba
p(\bar {p}) +p \to p(\bar {p}) + p +F
\ea
have a form [3]:
\ba
\biggl(\frac{s_1 d\sigma}{d s_1}\biggr)^{pp \to ppF} = \biggl(\frac{\alpha}{2\pi}\biggr)^2
\ln^2\biggl(\frac{s}{M_p^2}\biggr) \cdot f\biggl(\frac{s_1}{s}\biggr) \cdot \sigma_{tot}^{\gamma\gamma \to F}(s_1)
\biggl(1+O\biggl(\frac{1}{\ln s/{M_p^2}}\biggr)\biggr), \nn \\
f(z) = (2+z)^2 \ln\frac{1}{z} - 2(1-z)(3+z),
\ea
with $s_1$ - invariant mass square of the produced set of particles $F$.

%----------------------------------------------------
\section{QCD Processes. Check of $RRP$ vertex}
\label{SectionQCD1Processes}
%----------------------------------------------------

Using V.Gribov's prescription of Green function of exchanging gluon in process
$p(\bar p)(p_1) +p(p_2) \to jet(X_1)+jet(X_2)$
we put the matrix element in form
\ba
M^{pp \to j_1 j_2}=\frac{4\pi\alpha_s}{q^2}<X_1|J_{\mu}t^a|p_1><X_2|J_{\nu}t^a|p_2> \cdot
\frac{2}{s}p_2^{\mu}p_1^{\nu} = \frac{8\pi\alpha_s \cdot s}{q^2}\cdot \Phi_1^{a}\cdot \Phi_2^{a},
\ea
\ba
\Phi_1^a = \frac{1}{-s \alpha}<X_1|\vec {J} \vec {q} t^a|p_1>, \,\,\,\
\Phi_2^a = \frac{1}{s \beta}<X_2|\vec {J} \vec {q} t^a|p_2>,\,\,\,s=(p_1 +p_2)^2 >>M_p^2,
\ea
with $t^a$ - generator of color SU(N) group,
where we use the gauge conditions $q^{\mu}<X_{1,2}|J_{\mu}|p_{1,2}>=0$ and accept Sudakov
parametrization for 4-momentum of exchanged gluon $q=\alpha p_2 +\beta p_1 +q_{\bot}.$

Quantities $-s\alpha, s\beta$, can be interpret in terms of invariant mass squared of
fermionic jets created by initial protons.
$$
(p_1 -q)^2 \approx M_1^2 \approx -\vec{q}^2 - s\alpha; \,\,\,\,\,\,
(p_2 +q)^2 \approx M_2^2 \approx -\vec{q}^2 - s\beta.
$$
As was shown in papers [6] the phenomenon of "reggeization" of gluon Green function take
place in kinematics $s>> |q^2|$ - which consist in replacement of ordinary gluon to
reggeized gluon with the same quantum numbers except of "moving" gluon spin-it's Regge
trajectory. The matrix element of processes with reggeized
gluon exchange will acquire the Regge factor $R=\biggl(\frac{s}{s_0}\biggr)^{\alpha(q^2)}$, with
$\alpha(q^2)=1 - \alpha' \cdot \vec{q}^2$ - trajectory of gluon Regge pole (specified below).
\begin{figure}
\includegraphics[width=0.9\textwidth]{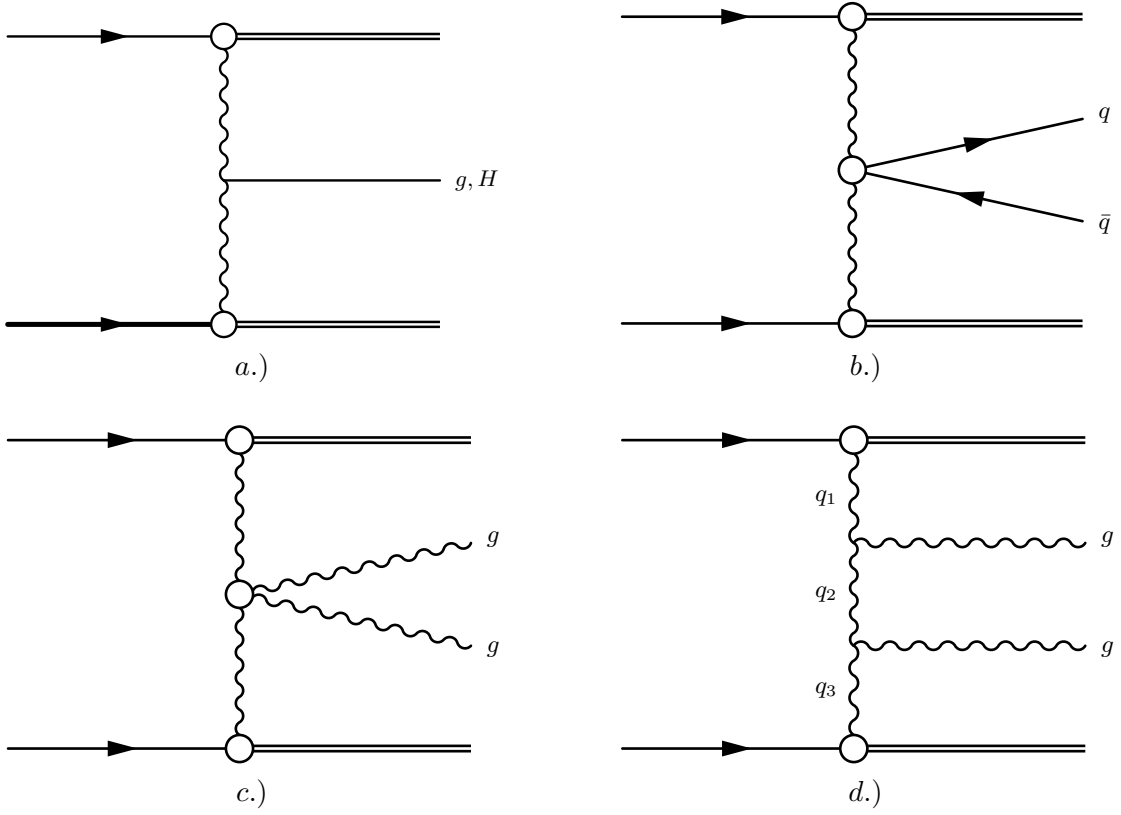}
\caption{Feynman diagrams of a single gluon jet and Higgs boson (a), quark-antiquark jets (b), \\
two gluon jets (c), two gluons separated by rapiditi gap (d).
\label{Fig2}}
\end{figure}
Matrix element of creation of an additional gluon have a form (see Fig 2a)
\ba
M^{PP \to j_1 j_2 g} = s \cdot \frac{(4\pi\alpha_s)^{3/2}}{q_1^2 q_2^2} \cdot
\frac{<X_1|\vec {J}\vec{q}_1t^a|p_1>}{-s \alpha_1} \cdot \frac{<X_2|\vec {J}\vec{q}_2t^b|p_2>}{s \beta_2} \cdot
f^{abc}C_{\mu}e_{\mu}^c(k), \nn \\
k=q_1 - q_2,
\ea
with
\ba
C_{\mu}=n_{-}^{\alpha}n_{+}^{\beta}\Gamma_{\alpha\beta\mu}, \,\,\,n_{-}=\frac{2p_1}{\sqrt {s}},\,\,\,n_{+}=\frac{2p_2}{\sqrt {s}},\,\,\,n_{+}^2=n_{-}^2=0,\,\,\,n_{+}n_{-}=2. \nn \\
C_{\mu}=2[(n_{-})_{\mu}(q_1^{+} +\frac{\vec {q}_1^2}{q_2^-})+ (n_{+})_{\mu}(q_2^{-} +\frac{\vec {q}_2^2}{q_1^+})-
(q_1 +q_2)_{\mu}],
\ea
effective vertex of conversion of two reggeized gluons to a real gluon, with properties
\ba
C_{\mu}(q1,q_2)(q_1 -q_2)_{\mu} =0; \,\,C_{\mu}^2 =\frac{16 \vec {q}_1^2 \vec {q}_2^2}{(-q_1^+ q_2^-)};\,\,
(q_1 -q_2)^2 = M_g^2 = -q_1^+ q_2^- -(\vec {q}_1-\vec {q}_2)^2.
\ea
In frameworks of fermion-jet model we replace the set of particles consisting in jet developed by
initial proton as on-mass shell proton, and, besides, modify the vertex of it's interaction with
(reggeized) gluon
\ba
<X_1|J_{\mu} t^a|p_1> = J_{\mu}^a = \bar {u}(p_1' +P_1)t^a \hat {V}_{\mu}u(p_1),\,\,\,
\hat {V}_{\mu} = \gamma_{\mu}-\frac{p_2^{\mu}}{q_1 p_2}\hat {q}_1.
\ea
This vertex function obey the gauge condition $J_1^{\mu}q_1^{\mu} =0.$
We have, besides
\ba
\frac{1}{s}J_{\mu}^a p_2^{\mu} =\frac{1}{-s\alpha_1}\vec{J^a}\vec{q}_1^a; \nn \\
\int d(s\alpha_1)d\gamma_1 \sum \biggl(\frac{\vec{J}_{\mu}^a p_2^{\mu}}{s}\biggr)
\biggl(\frac{\vec{J}_{\nu}^b p_2^{\nu}}{s}\biggr) = \frac{1}{2}\delta_{ab}\frac{2\vec{q}_1^2}{\bar {M}^2+\vec{q}_1^2}
\ea
with $d\gamma_1$ - phase volume of proton jet defined in (29) and $\bar M$ - the average value of  invariant mass of proton jet.

\begin{figure}
\includegraphics[width=0.9\textwidth]{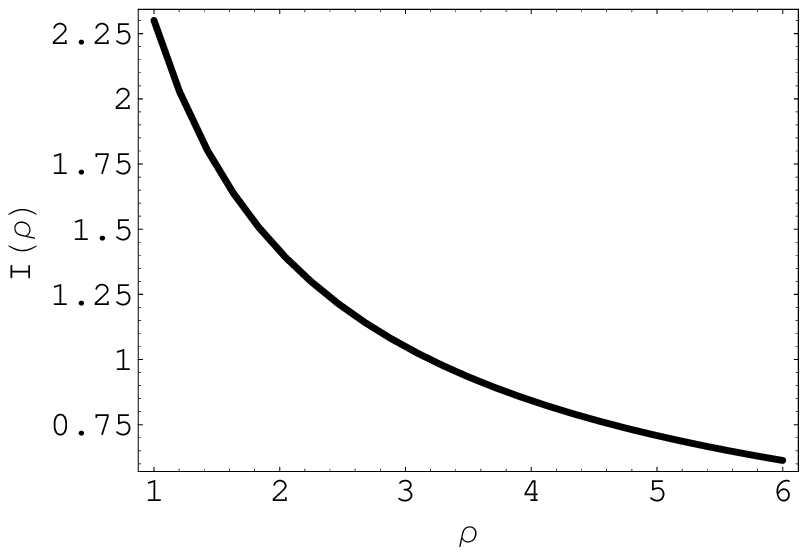}
\caption{$I(\rho)$ (see (35)) for process $pp \to j_1j_2j_g$.
\label{Fig3}}
\end{figure}

For matrix element squared overaged on final state we obtain in fermion-jet model:
\ba
\int d(s\beta_2)d\gamma_2 \int d(s\alpha_1)d\gamma_1 \sum|M|^2 = \frac{s^2 \cdot 2^6 \pi^3 \alpha_s^3}{\bar{q}_1^2 \bar{q}_2^2} \cdot
\frac{N(N^2-1)}{M_g^2+(\vec{q}_1 -\vec{q}_2)^2} \cdot
\frac{\vec{q}_1^2}{M^2+\vec{q}_1^2}\frac{\vec{q}_2^2}{M^2+\vec{q}_2^2}
\ea
Considering the phase volume of process $p(\bar p)p \to j_1 j_2 F$ we
introduce two auxiliary variables - the 4-momenta of the exchanged gluons
\ba
\int d^4q_1 d^4q_2 \delta^4(p_1 -q_1 -p_1' -P_1)\delta(q_2 +p_2 -p_1' -P_2)=1.
\ea
So we have
\ba
d\Gamma_3 = (2\pi)^{-2}d^4q_1 d^4q_2 d\gamma_1 d\gamma_2 d\gamma_j,\,\,\,
d\gamma_1=\frac{d^3p_1'}{2\varepsilon_1'}\Pi_{P_1} \frac{d^3 r_i}{2\varepsilon_i (2\pi)^3}\delta^4(p_1-q_1-p_1'-P_1), \,\,
P_1 =\sum r_i, \nn \\
d\gamma_2=\frac{d^3p_2'}{2\varepsilon_2'}\Pi_{P_2} \frac{d^3 v_i}{2\varepsilon_i (2\pi)^3}\delta^4(p_2+q_2-p_2'-P_2),\,\,\,\,
P_2 =\sum v_i, \nn \\
d\gamma_j=\Pi_{P_j} \frac{d^3 l_i}{2\varepsilon_i (2\pi)^3}\delta^4(q_1-q_2-P_j),\,\,\,\,
P_j =\sum l_i. \qquad
\ea
Using
\ba
d^4q_1 d^4q_2 =\frac{s}{2}d\alpha_1 d\beta_1 d^2q_1 \cdot \frac{s}{2} d\alpha_2 d\beta_2 d^2q_2 =
\frac{\pi^2}{4s}d(s \alpha_1)d(s\beta_2)\frac{d\beta_1}{\beta_1}d(s\alpha_2\beta_1)\frac{d^2q_1 d^2q_2}{\pi^2}
\ea
we write down $d\Gamma_3$ in form
\ba
d\Gamma_3 = \frac{1}{2^7 \pi^3 s}\frac{d\beta_1}{\beta_1} \cdot d\Phi_1 \cdot d\Phi_2 \cdot d\Phi_g
\frac{d^2q_1 d^2q_2}{\pi^2}; \,\,\,d\Phi_{1,2}= dM_{1,2}^2 d\gamma_{1,2};\,\,\,d\Phi_g = dM_g^2 d\gamma_j.
\ea
In fermion - jet model approximation we obtain
\ba
d\Gamma^{(3)}=(2\pi)^{-5} \cdot \frac{\pi^2}{4 s}\cdot \frac{d\beta_1}{\beta_1}\cdot \frac{d^2q_1 d^2q_2}{\pi^2}
\ea
For cross section of process $pp \to j_1 j_2 j_g$ we obtain
\ba
d\sigma = \frac{\alpha_s^3}{16M_g^2}N(N^2-1)R_2 dL_1 \cdot I(\rho),\,\,\,dL_1 = \frac{d\beta_1}{\beta_1}
\ea
\ba
R_2=\biggl(\frac{s_1}{s_0}\biggr)^{2(\alpha(\bar{q}_1^2)-1)} \biggl(\frac{s_2}{s_0}\biggr)^{2(\alpha(\bar{q}_0^2)-1)}
\approx \biggl(\sqrt{s} (GeV)\biggr)^{-4\frac{\alpha_s}{\pi} \cdot {\vec{q}^2}(GeV^2)},
\ea
\ba
I(\rho) = \int\limits_0^{\infty}\int\limits_0^{\infty}\frac{dx_1 d_2}{(x_1+\rho)(x_2+\rho)\sqrt{(1+x_1+x_2)^2-4x_1x_2}},
\,\,\,\,\rho=\frac{\bar{M}^2}{M_g^2}.
\ea
Function $I(\rho)$ is tabulated in Fig.~\ref{Fig3}.
%
%----------------------------------------------------
\section{QCD Processes. Check of $RRPP$ vertex}
\label{SectionQCD2Processes}
%----------------------------------------------------
Let now consider process of creation of two gluons not separated by rapidity gap
\ba
p(p_1)+p(p_2) \to jet_1(X_1)+jet_2(X_2)+g(k_1)+g(k_2).
\ea
For the case of production of two particles $RR \to a(k_1)+b(k_2)$ (no rapidity gap between
$a,b$ implied) we obtain for phase volume
\ba
d\Gamma_4 = \frac{1}{2^{11}\pi^5}\frac{dx}{x \bar x} \cdot \frac{d^2q_1 d^2q_2 d^2k_1}{\pi^3}
\ea
with $k_i = b_ip_1 +a_ip_2 +k_{\bot}, \,\,\,x=\frac{b_1}{\beta_1};\,\,y=\frac{a_1}{-\alpha_2}$. \\
\begin{table}
\begin{tabular}{|c|c|c|c|c|}
\hline
$\sqrt {s}(GeV) \backslash  q^2 (GeV^2)$ & 0.5 & 1 & 3 & 5 \\

\hline
1000 & 0.2512 & 8.0631 & 0.00025 & 10$^{-6}$ \\
\hline
7000 & 0.17021 & 0.0289 & 0.000024 & 2.041 $\cdot$ 10$^{-8}$ \\
\hline
14000 & 0.1482 & 0.02196 & 0.0000106 & 5.102 $\cdot$ 10$^{-9}$ \\

\hline
\end{tabular}
\caption{Estimation of gluon reggeization factor $R_2$}
\label{TableR}
\end{table}

\begin{figure}
\includegraphics[width=0.9\textwidth]{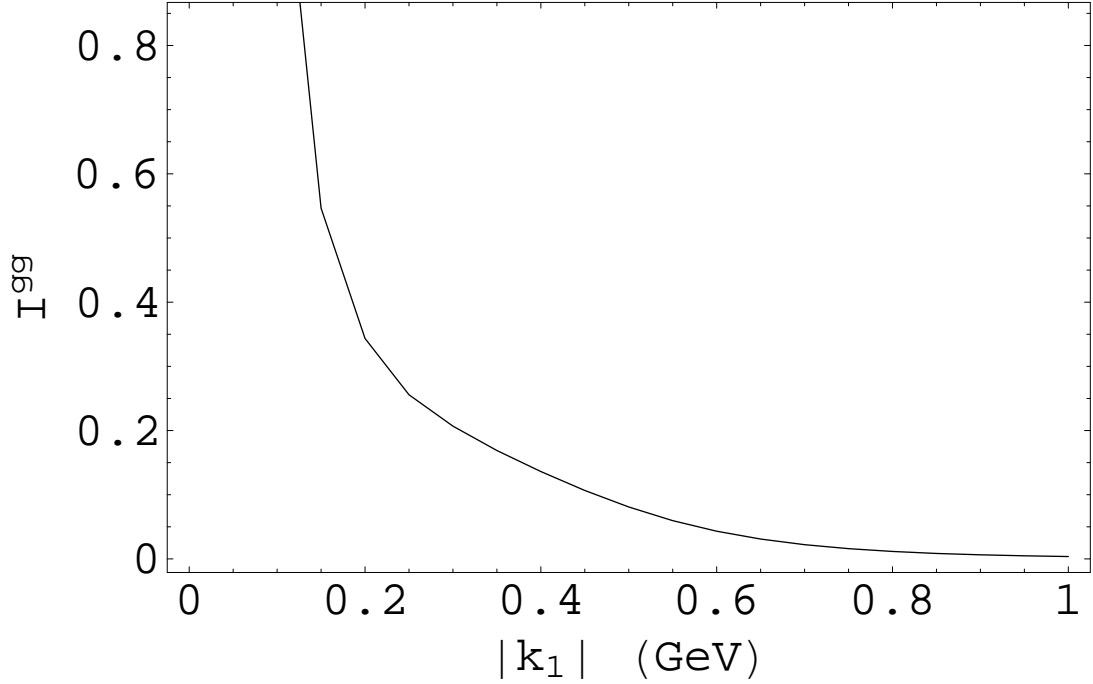}
\caption{Value $I_{gg}$ (defined in (\ref{IggDef})) as a function of transverse momentum modulus
$|\vec k_1|$ of one of the gluons in the produced gluon pair
in the case of $M_1=M_2=1\GeV$, $x=0.2$ and $y=0.3$.
\label{FigIgg}}
\end{figure}

Differential cross section of pair of particles $ab$ ($ab = gg, q\bar {q}$)
production in fermion-jet model can be written as
\ba
d\sigma^{pp \to j_1 j_2 ab} = \frac{\alpha_s^4}{2^6 \pi} d L \cdot R_2 \cdot \frac{dx}{x(1-x)} \cdot \frac{d^2k_1}{\pi}
\frac{d^2q_1 d^2q_2}{\pi^2}\Phi^{ab}, \,\,\,\,\,\Phi^{ab}=\frac{\sum|M^{RRab}|^2}{\bar {q}_1^2 \bar {q}_2^2}.
\ea
The explicit expression for $\Phi^{gg}$ is ~\cite{Antonov:2004hh,Fadin:1996nw} ($M^{RRgg}=M^{RRPP}$):

It was obtained \cite{Lipatov:1995pn,Antonov:2004hh,FKL,Fadin:1989kf,Fadin:1996nw}:
\ba
\sum|M^{RRPP}|^2&=&G_1(a^{\nu_1\nu_2}(k_1,k_2))^2+G_2\Omega_{\sigma\sigma'}(k_1)\Omega_{\rho\rho'}(k_2)
a^{\sigma\rho}(k_1,k_2)a^{\rho'\sigma'}(k_2,k_1)+\nn\\
&+&(k_1\leftrightarrow k_2),
\ea
where
\ba
G_1&=&(f_{d_1d_2r}f_{cdr})^2=N^2(N^2-1); \nn \\
G_2&=&f_{d_1d_2r}f_{cdr}f_{d_2cr}f_{d_1dr}=-\frac{1}{2}N^2(N^2-1),
\ea
the projection operators
\ba
\Omega_{\sigma\sigma'}(k)=-g_{\sigma\sigma'}^\bot-\frac{2}{\vec{k}^2}k_{\sigma\bot}k_{\sigma'\bot},
\ea
and
\ba
a^{\nu_1\nu_2}(k_1,k_2)=4\biggl[\frac{1}{t}q_\bot^{\nu_1}q_\bot^{\nu_2}-
\frac{1}{\chi}q_\bot^{\nu_1}(k_1-\frac{x}{\bar{x}}k_2)^{\nu_2}+
\frac{1}{\chi}q_\bot^{\nu_2}(k_2-\frac{\bar{y}}{y}k_2)^{\nu_1}-
\frac{x\vec{q}_2^2}{\chi\vec{k}_1^2}k_1^{\nu_1}k_1^{\nu_2}- \nn \\
\frac{\bar{y}\vec{q}_1^2}{\chi\vec{k}_2^2}k_2^{\nu_1}k_2^{\nu_2}-
\frac{1}{\chi}(1+\frac{tx}{\bar{x}\vec{k}_1^2})k_1^{\nu_1}k_2^{\nu_2}+
\frac{1}{\chi}k_1^{\nu_1}k_2^{\nu_2}-2Dg_\bot^{\nu_1\nu_2}\biggr],
\ea
with
\ba
D=1+\frac{t}{\chi}+\frac{\bar{x}\vec{k}_1^2}{tx}+\frac{1}{\chi}[\frac{\bar{x}}{x}\vec{k}_1^2-
\frac{x}{\bar{x}}\vec{k}_2^2]+\frac{\vec{q}_1^2}{\chi}\bar{y}+\frac{\vec{q}_2^2}{\chi}x.
\ea
Using the relations
\ba
s a_i b_i=\vec{k}_i^2 + m^2; \nn \\
q=q_1-k_1=q_2+k_2; t=q^2;\chi=(k_1+k_2)^2; \nn \\
\ea
\ba
t=-(\vec{q}_1-\vec{k}_1)^2-\frac{\bar{x}}{x}\vec{k}_1^2; \chi=\frac{1}{x\bar{x}}(\bar{x}\vec{k}_1-x\vec{k}_2)^2,
\ea
one can be convinced that the gauge conditions
\ba
\left.D\right|_{\vec{q}_1\to 0}=\left.D\right|_{\vec{q}_2\to 0}=0; \qquad
\left.a^{\nu_1\nu_2}(k_1,k_2)\right|_{\vec{q}_1\to 0}=0; \nn \\
\left.a^{\nu_1\nu_2}(k_1,k_2)\right|_{\vec{q}_2\to 0}=0,
\ea
are fulfilled. \\
Due to gauge properties of $a^{\nu_1\nu_2}$ the quantity $\Phi^{gg}$ is finite as
$\vec{q}_1,\vec{q}_2 \to 0$, which provides the convergence of the quantity
\ba
I^{gg}(\vec{k}_1)=\int\frac{d^2\vec{q}_1d^2\vec{q}_2 (/\pi^2)}{(\vec {q}_1^2+\bar M^2)(\vec {q}_2^2+\bar M^2)}\Phi^{gg}.
\label{IggDef}
\ea
This quantity is presented in Fig.~\ref{FigIgg}.
\begin{figure}
\includegraphics[width=0.9\textwidth]{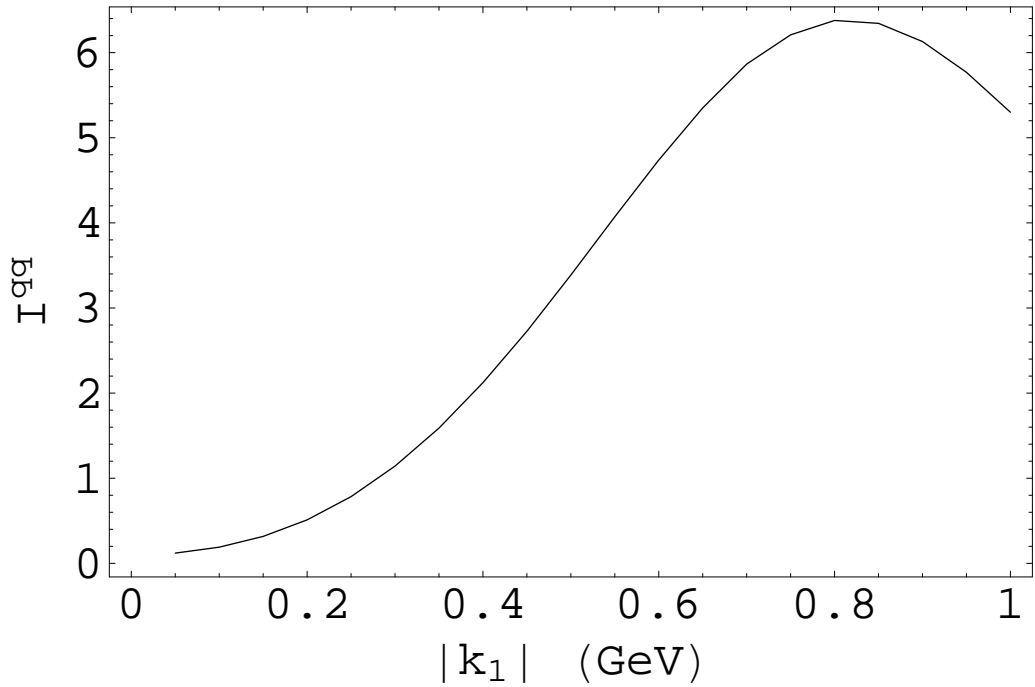}
\caption{Value $I_{qq}$ (defined in (\ref{IqqDef})) as a function of transverse momentum modulus
$|\vec k_1|$ of quark in the produced quark pair
in the case of $M_1=M_2=1\GeV$, $x=0.2$ and $y=0.3$.
\label{FigIqq}}
\end{figure}
%
%\begin{figure}
%\includegraphics[width=0.9\textwidth]{Fig3.eps}
%\caption{Feynman diagrams of a single gluon jet (a), quark-antiquark jets (b),two gluon jets (c)
%\label{Fig2}}
%\end{figure}
%
%\begin{figure}
%\includegraphics[width=0.9\textwidth]{Ig.eps}
%\caption{Value $I_g$ (defined in (\ref{IgDef})) as a function of the produced gluon jet mass $M$ (in GeV) in the
%case of $M_1=M_2=1\GeV$.
%\label{FigIg}}
%\end{figure}
%
\section{Azimuthal correlation in process of two gluon jets creation, separated by rapidity gap}
\label{Section5}

Consider now process of two gluon production separated by rapidity gap.
Corresponding matrix element contains 3 gluon Regge factors
$R_3 = \biggl(\frac{s_1}{s_0}\biggr)^{\alpha(q_1)} \biggl(\frac{s_2}{s_0}\biggr)^{\alpha(q_2)}  \biggl(\frac{s_3}{s_0}\biggr)^{\alpha(q_3)}$ with momenta of exchanged gluons
$q_i = \alpha_i p_2 +\beta_i p_1 +q_{1\bot}$ and $s_1 \approx -s \alpha_2,\,\,s_2 \approx -s \alpha_3 \beta_1,\,\,\,
s_3 = s\beta_2, \,\,\,1 >> \beta_1 >> \beta_2 >>\beta_3,\,\,\,1>>\alpha_3 >>\alpha_2>>\alpha_1.$
\ba
s_1s_2s_3 = [M_1^2+(\bar {q}_1 -\bar {q}_2)^2][M_2 +(\bar {q}_2 -\bar {q}_3)^2]\cdot s,
\ea
and $M_1^2, M_2^2$ - the invariant mass squared of created gluon jets.

Matrix element have a form
\ba
M^{pp \to j_1j_2 j_{g_1}j_{g_2}} = \frac{s(4\pi\alpha_s)^2}{2 q_1^2 q_2^2 q_3^2}
\frac{<X_1|\bar {J}\bar {q}_1 t^a|p_1>}{(-s \alpha_1)} \cdot \frac{<X_2|\bar {J}\bar {q}_2 t^b|p_2>}{s \beta_3} \cdot \nn \\
f_{adc_1}f_{bdc_2}C^{\mu}(q_1, q_2)e_{\mu}^{c_1} \cdot C^{\lambda}(q_2, q_3)e_{\lambda}^{c_2}.
\ea
Phase volume of process $pp \to j_1j_2 j_{g_1}j_{g_2}$ can be written in form (in fermion-jet model)
\ba
d\Gamma_4 = (2\pi)^{-8}\frac{\pi^3}{8s} \frac{d\beta_1}{\beta_1} \frac{d\beta_2}{\beta_2} \frac{d^2q_1 d^2q_2 d^2q_3}{\pi^3}.
\ea
For cross section we obtain
\ba
d\sigma^{pp \to 4j} = \frac{4\alpha_s^4}{\pi} dL_1 \Delta Y \frac{d\bar{q}_1^2 d\bar{q}_3^2 (d\varphi/(2\pi)) (d^2q_2/(\pi))R_3 N^2(N^2-1)}{(\bar {q}_1^2 +\bar {M}^2)(\bar {q}_3^2 +\bar {M}^2)(M_{g_1}^2 +(\bar {q}_1 -\bar {q}_2)^2)
(M_{g_2}^2 +(\bar {q}_2 -\bar {q}_3)^2)},
\ea
with
$M_{p_1}^2 =-s\alpha_1;\,\,\,M_{p_2}^2 =-s\beta_3;\,\,\,M_{j_1}^2 =-s\alpha_2 \beta_1;\,\,\,M_{j_1}^2 =-s\alpha_3 \beta_2.$
and
$L_1=\ln\frac{s\beta_1}{M^2},\,\,\Delta Y$ - rapidity gap of gluon jets $\Delta Y =\ln\frac{\beta_1}{\beta_2}$.

\begin{figure}
\includegraphics[width=0.9\textwidth]{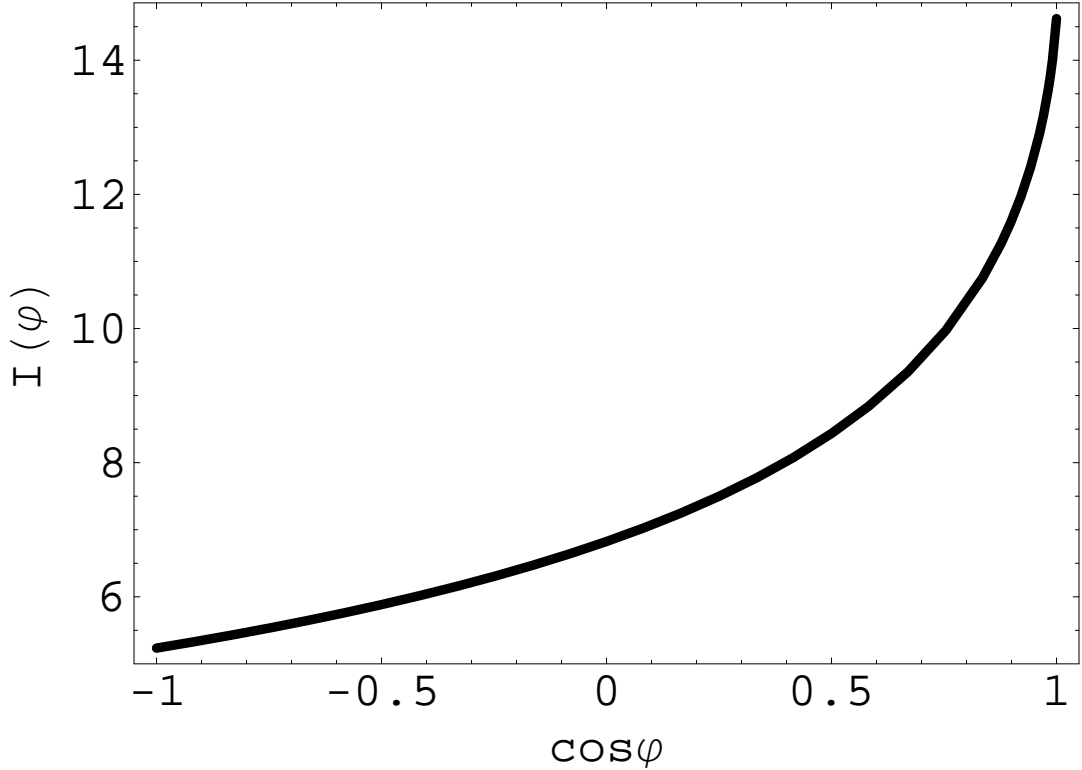}
\caption{Azimuthal correlation of two gluon jet production separated by rapidity gap (see (53)).
\label{Fig6}}
\end{figure}

For azimuthal correlation we obtain (performing integration on $d^2q_2$ and puting $M_{g_1}^2=M_{g_2}^2=M_g^2$)
\ba
\frac{2\pi d\sigma^{pp \to j_1j_2 j_{g_1}j_{g_2}}}{dY d\varphi} = L_1 \sigma_0 \cdot F(\varphi), \,\,\,
L_1 =\ln\frac{s\beta_1}{M^2}, \,\,\,Y=\ln\frac{\beta_1}{\beta_2},\,\,\,\sigma_0 = \frac{4\alpha_s^4}{\pi M_g^2}N^2(N^2-1).
\ea
\ba
F(\varphi) = \int\limits_0^{\infty}\frac{dx_1}{x_1+\rho}\int\limits_0^{\infty}\frac{dx_2}{x_2 +\rho}\psi(z),\,\,\,
z=x_1 +x_2-2\sqrt{x_1 x_2}\cos\varphi, \nn \\ \psi(z)=\frac{2}{\sqrt{z(4+z)}}\ln\frac{\sqrt{4+z}+\sqrt{z}}{\sqrt{4+z}-\sqrt{z}},\,\,\,\,\,\rho=\frac{\bar {M}^2}{M_g^2}.
\ea
Function $F(\varphi)$ is presented in Fig.~\ref{Fig6}.
%

%----------------------------------------------------
\section{The Process $pp\to jj q\bar{q}$. Check of $RRqq$ vertex}
\label{Sectionjjqq}
%----------------------------------------------------

The matrix element of the subprocess of conversion of two reggeized gluons to the quark-anti-quark
pair (see Fig.~\ref{Fig2}, c)
\ba
R(a,q_1)+R(b,-q_2) \to q(k_1)+\bar{q}(k_2),
\ea
is described by two different mechanisms: direct interaction and production of gluon with
its subsequent conversion into the quark pair ~\cite{Fadin:1996tb}
\ba
M^{q\bar{q}}=\bar{u}(k_1)[A t_at_b-B t_bt_a]v(k_2),
\ea
with $t_a$ being the generator of the color $SU(N)$ group in the fermion representation,
\ba
\sum (t_a^2)^2=I\frac{N^2-1}{2N}; \sum Tr t_at_bt_at_b=-\frac{N^2-1}{4N}; \sum Tr t_at_at_bt_b=\frac{(N^2-1)^2}{4N},
\ea
and \cite{Fadin:1989kf,Fadin:1996nw}
\ba
A&=&\gamma^-\frac{\hat{q}_1-\hat{k}_1-m}{(q_1-k_1)^2-m^2}\gamma^+-\frac{1}{q^2}\hat{C}; \qquad \gamma_\pm = n_\pm^\mu \gamma_\mu,
\nn \\
B&=&\gamma^+\frac{\hat{q}_1-\hat{k}_2-m}{(q_1-k_2)^2-m^2}\gamma^--\frac{1}{q^2}\hat{C}, \qquad q=k_1+k_2,
\ea
where $m$ is the quark mass and 4-vector $C_\mu$ describing the conversion of two reggeized gluons into
the ordinary gluon which was given above (23).
The gauge properties of $M^{q\bar{q}}$, i.e. turning it to zero in the limit $\vec{q}_1\to 0$ as well as
in the limit $\vec{q}_1\to 0$ can be seen explicitly.
These properties provide convergence of the relevant integrals on $\vec{q}_{1,2}$.
We obtain
\ba
\Phi^{q\bar{q}}=\frac{4M_{q \bar q}^4}{\vec{q}_1^2\vec{q}_2^2}[N_1(S_A+S_B)-2N_2S_{AB}], \,\,\,\,
N_1=\frac{(N^2-1)^2}{4 N};\,\,N_2=-\frac{(N^2-1)}{4 N},
\ea
\ba
M_{q \bar q}^2 =\frac{1}{x \bar x}[m^2+(\vec{k}_1+x(\vec{q}_2-\vec{q}_1))^2]
\ea
with
\ba
S_A=\frac{1}{4}Sp(\hat{k}_1+m)A(\hat{k}_2-m)\tilde{A}; \nn \\
S_B=\frac{1}{4}Sp(\hat{k}_1+m)B(\hat{k}_2-m)\tilde{B}; \nn \\
S_{AB}=\frac{1}{4}Sp(\hat{k}_1+m)A(\hat{k}_2-m)\tilde{B}.
\ea
Note that the value $\Phi^{q\bar{q}}$ is finite in both the limits $\vec{q}_1\to 0$ and $\vec{q}_2\to 0$.

The result of numerical integration of the quantity
\ba
I^{qq}=\int\frac{d^2\vec{q}_1d^2\vec{q}_2}{\pi^2} \frac{M^4 \cdot \Phi^{q\bar{q}}}{(M^2+\vec{q}_1^2)(M^2+\vec{q}_2^2)}
\label{IqqDef}
\ea
is presented in Fig.~\ref{FigIqq}. \\
Cross section of process $pp \to j_1j_2 q \bar{q}$ is:
\ba
d\sigma^{pp \to j_1j_2 q \bar{q}} = \frac{\alpha_s^4}{2^6 \pi M^2}\frac{d^2k_1}{\pi M^2}\frac{dx}{x(1-x)}I^{qq} \cdot
\frac{d\beta_1}{\beta_1} \cdot \bar{R}_2.
\ea
\section{Higgs boson production}
\label{Higgs}

Higgs boson, we assume, to be produced by collision of two reggeized gluons through intermediate
state of heavy top quark - antiquark state. By analogy with QED case we have for matrix element (see above)
\ba
M^{pp \to j_1 j_2 H} = \frac{\pi\alpha_s}{q_1^2 q_2^2} \cdot \frac{2\alpha_s N g_H}{\pi m_t}I_s \cdot
\vec{q}_1\vec{q}_2 \frac{<X_1|\vec{J}\vec{q}_1 t^a|p_1>}{-s\alpha_1} \cdot
\frac{<X_2|\vec{J}\vec{q}_2 t^a|p_2>}{s\beta_2}.
\ea
For differential cross section we obtain
\ba
\frac{d\sigma}{dL} = \sigma_0 \gamma, \,\,\,\sigma_0 = \frac{\alpha_s^4 N^2(N^2-1)}{2^9 \pi^3 m_t^2} |I_s|^2, \,\,
\gamma = \int\frac{d^2q_1 d^2q_2/{\pi^2}}{(\bar{M}^2+\vec{q}_1^2)(\bar{M}^2+\vec{q}_2^2)}R_2
\ea
with $I_s \biggl(|\frac{M_H}{m_t}|^2, \frac{\vec{q}_1^2}{m_t^2}, \frac{\vec{q}_2^2}{m_t^2}\biggr) \approx I_s(0,0,0) = \frac{1}{3}$. \\
Quantity $\gamma$ turns out to be small $\gamma \sim 10^-2$ for $\frac{\alpha_s}{\pi} = 0.1 \,\,\,
\sqrt{s} \sim 10^3 \div 10^4 \,\,GeV$. So the differential on Higgs boson rapidity cross section $\sigma_0$
is rather small $\gamma \cdot \sigma_0 \sim 1 \,fb, \,\,\,\,\sqrt{s} = 14000 \,\,GeV$.

\section{Discussion}

The channel of peripherical processes with creation of some state in $s_0$ - called pionization region in
proton (anti-proton) - proton collisions at high energies are considered. We suppose the initial state proton
(antiproton) to develop the protonic (anti-protonic) jets resulting from interaction of initial proton
(anti-proton) with the reggeized (coloured) gluon.

The fact of "reggeization" of gluon - the replacement of Green function of the exchanged gluon by the
excahange of a Regge - pole with gluon quantum numbers except of a spin, which is replaced by gluon Regge
trajectory
\ba
\alpha(q^2) = 1+ \frac{\alpha_s q^2}{\pi}\int\frac{d^2k}{k^2(q-k)^2}
\ea
was proven in papers ~\cite{Kuraev:1976ge}. Gluon Regge trajectory suffer from infrared singularities,
which can be regularized by  introducing the fictitious gluon mass $m$: $k^2 \to k^2+m^2, \,\,\,
(q-k)^2 \to (q-k)^2 +m^2$. In papers of Lipatov and Balitskii ~\cite{Balitsky1978} was shown that the singular dependence
on gluon mass disappears a experimental set -up with emission of (arbitrary) "real" gluons, as well,
posessing mass. It means taken into account the inelastic processes with creation of so called
"mini-jets" in multi -Regge kinematics. In this way the expression of cross section in terms of Pomeron -pole
exchange was developed. The corresponding forward scattering amplitude was shown ~\cite{Kuraev:1976ge} to obey the so-called
BFKL equation.

The statement of gluonic "mini-jets" is some problematic up to now. Really by the common knowledge the
gluon color must manifest itself in developing the hadronic jet, consisting from pions.

In this paper we use the theoretical result about existence of gluon trajectory and use the phenomenologic
approach in describing it's form ~\cite{Kaidalov:2001db}
\ba
\alpha(q) = 1-\frac{\alpha_s}{\pi} \frac{\vec{q}^2}{q_0^2}C, \,\,\,q^2 =-\vec{q}^2, <0, \,\,\,q_0^2 \approx 1 \,\,GeV^2, \,\,\,C \approx 1.
\ea
Besides we use some simplified version of GPD describing the interaction of reggeized gluon with proton (anti-proton)
and converging it to proton-jet.

Main feature of this "fermion-jet" model consist in absence of evolution effects, which is the essential part
of Generalized Parton Distribution (GPD) approach. Applying  simultaneously BFKL and evolution mechanisms
seems to be suffer from the double - counting.  \\
Regge-factor writtenin form (two and three successive reggeon (R) exchange)
\ba
R_2 = \biggl(\frac{s_1}{s_0}\biggr)^{2(\alpha(q_1)-1)} \biggl(\frac{s_2}{s_0}\biggr)^{2(\alpha(q_2)-1)};\,\,\,\,
R_3 = \biggl(\frac{s_1}{s_0}\biggr)^{2(\alpha(q_1)-1)} \biggl(\frac{s_2}{s_0}\biggr)^{2(\alpha(q_2)-1)} \biggl(\frac{s_3}{s_0}\biggr)^{2(\alpha(q_3)-1)}
\ea
$s_1s_2 = M_F^2 s;\,\,\,s_1 s_2 s_3 =M_{1F}^2M_{2F}^2s$ \\
with $M_F^2$ -invariant mass created in pionization regions, turns out to be a rather effective suppression factor.
In papers ~\cite{Hager} the of gluon reggeization effects was omitted.

In paper ~\cite{Khoze} the channel $pp \to ppH$ aof Higgs boson production was investigated. Here at least the
exchange by two (parallel) (reggeized) gluons must be applied to provide colorless $ppH$ final state. \\
Introducing the Sudakov - type formfactors as well seems to be illegetimatle.

Cross section of Higgs boson production in our approach is rather small $d\sigma/dL \sim 1\,\,fb$, but can
be measured at LHC. This result is in agreement with ones obtained in ~\cite{Khoze,Georgi}.
Tests of effective Regge action theory developed in paper ~\cite{Antonov:2004hh} provide by investigation of processes of
creation of a single gluonic jet, and production of two gluons and quark-anti-quark pair in pionization
region (without rapidity gap between created gluon or quarks). In these experiments the form $RRP, RRPP, RRq\bar q$ -
vertices of effective Regge action can be tested.

Measuring the azimuthal correlation in process of production of two gluon jets, separated by rapidity gap
as a test of theory prediction for multi-regge kinematics as well can be realized at RHIC or LHC.
\acknowledgements

We are grateful to Lev~Lipatov, Vladimir~Saleev and Ahmed~Ali for discussions and critical remarks.
We acknowledge the support of RFBR, grant no. 10-02-01295a.
This work was also supported by grant HLP-2010-06 of the Heisenberg--Landau program, and
JINR--Belorusia--2010 grant.

\end{document}